\documentclass{amsart}
\pdfoutput=1
\usepackage{graphicx}
\usepackage{url}
\usepackage{booktabs}
\usepackage{amsaddr}
\usepackage{amsbsy}
\usepackage{amsfonts}
\usepackage{amssymb}
\usepackage{amsmath}
\usepackage{natbib}
\newcommand{\bds}[1]{\boldsymbol{#1}}
\newcommand{\bse}{\boldsymbol{\eta}}

\newcommand{\bsG}{\boldsymbol{G}}
\newcommand{\bsA}{\boldsymbol{A}}

\newcommand{\bsR}{\boldsymbol{R}}

\newcommand{\bsS}{\boldsymbol{S}}

\newcommand{\bsX}{\boldsymbol{X}}
\newcommand{\bsV}{\boldsymbol{V}}

\newcommand{\bss}{\boldsymbol{s}}

\newcommand{\bsx}{\boldsymbol{x}}

\newcommand{\bsy}{\boldsymbol{y}}
\newcommand{\bsv}{\boldsymbol{v}}

\newcommand{\bstheta}{\boldsymbol{\theta}}

\newcommand{\moparm}{\boldsymbol{\Psi}}
\newcommand{\bsSigma}{\boldsymbol{\Sigma}}

\newcommand{\bsbeta}{\boldsymbol{\beta}}
\newcommand{\bsGamma}{\boldsymbol{\Gamma}}
\newcommand{\te}[1]{\textrm{#1}}
\newcommand{\der}[1]{\frac{\partial}{\partial #1}}

\newcommand{\ud}{\mathrm{d}}
\newcommand{\Cov}{{\mbox{Cov}}}

\usepackage{graphicx}

\usepackage{amssymb}

\begin{document}

\author[Karl, Yang, and Lohr]{Andrew T. Karl \and Yan Yang}
\address{Arizona State University}
\author{Sharon L. Lohr}
\address{Westat}
\title[MLE of Multiple Membership Linear Mixed Models]{Efficient Maximum Likelihood Estimation of Multiple Membership Linear Mixed Models, with an Application to Educational Value-Added Assessments}

\begin{abstract}
The generalized persistence (GP) model, developed 
in the context of estimating
``value added'' by individual teachers to their students' current
and future test scores, is one of the most flexible value-added models in the literature.  
Although developed in the educational setting, the GP model can potentially be applied to any structure where each sequential response
of a lower-level unit may be associated with a different higher-level unit, and the effects of the higher-level units may persist over time.
The flexibility of the GP model, however,
and its multiple membership random effects structure
lead to computational challenges that have limited the model's availability.  
We develop an EM algorithm to compute maximum likelihood
estimates efficiently for the GP model, making use of
the sparse structure of the random effects and error covariance matrices. The algorithm is
implemented in the package GPvam in R statistical software.
We give examples of the computations and illustrate the
gains in computational efficiency achieved by our estimation procedure.
\end{abstract}

\maketitle

\noindent\textit{Revision note (October 2026).} With GPT6 Astra's assistance, this version makes minor improvements and
corrections to the mathematical exposition and notation, and adds a brief subsection on REML estimation.
Numerical results are retained from the original analysis and have not been recomputed.
\par\medskip

\section*{NOTICE}
This is the author's version of a work that was accepted for publication in \textit{Computational Statistics \& Data Analysis}. Changes resulting from the publishing process, such as peer review, editing, corrections, structural formatting, and other quality control mechanisms may not be reflected in this document. Changes may have been made to this work since it was submitted for publication. A definitive version was subsequently published in \textit{Computational Statistics \& Data Analysis}, [VOL59, March, (2013)] DOI:10.1016/j.csda.2012.10.004

\section{Introduction}\label{sec:intro}
Multilevel mixed models are popular for describing data with complex dependence structure.
The units on which primary measurements are taken (usually those at the lowest level) each belong to
one or more units at higher levels.
In a nested (hierarchical) two-level model, each unit at the lowest level belongs to exactly one higher-level unit.
In a multiple membership structure \citep{browne01}, a lower-level unit may be associated with multiple higher-level units.
This structure is common with non-static populations, and we study multiple membership models in which
a lower-level unit is sequentially associated with different higher-level units.
Thus, a child in foster care may live with multiple families; a patient may see multiple doctors;
a deer may visit multiple salt licks; a worker may have multiple employers;
a person may attend multiple therapy groups;
a student may have multiple teachers.
\citet{fielding} describe multiple membership models and give examples of their use.
The multiple membership structure induces a complex dependence structure in the data.
Lower-level units are correlated whenever they share any higher-level unit, so the covariance matrix
will not have a block diagonal structure as in the nested model.

The complex covariance structure of multiple membership mixed models makes computations challenging, particularly
with large data sets.
Computational methods that have been developed for nested hierarchical models and other special cases of linear mixed models often
will not work. In this paper we develop an EM algorithm to compute maximum likelihood estimates for a class of longitudinal multiple membership
models that are applicable in many settings.
In the class of models considered, lower-level units are associated with multiple higher-level units in sequence,
and a response is recorded on a lower-level unit after the association with each higher-level unit.
If the population contains a large number of higher-level units, and the number of lower-level units
associated with each higher-level unit is bounded,
the covariance matrix will be sparse.
The algorithm exploits sparseness of the covariance matrix to speed computations.
This sparseness is achieved in many multiple membership settings since, for example, there are upper bounds on the number
of patients a doctor can see or the number of students in a teacher's class.

The application motivating this research comes from value-added models (VAMs) in educational evaluation.
A VAM score for a teacher is intended to estimate the ``value added'' by that teacher
to students' knowledge---how much more (or less) students' scores changed under that teacher
than they would be expected to change under an ``average'' teacher---by apportioning students' progress on standardized tests
to the teachers or schools that have taught those students.
\citet{braun} describe some of the potential uses of VAMs and discuss issues associated with using them to evaluate teachers and schools.

While a variety of different models are used (see \citet{lohr2012} for a review of common VAMs), in this paper we primarily consider the generalized persistence (GP) model developed by \citet{mariano10}, one of the most flexible models in the literature.
In the GP model, each student  is followed over  $T$ grades with a different
teacher in each grade, and receives a score on a standardized test at the end of each grade.
Each student therefore ``belongs'' to up to $T$ different teachers, resulting in a multiple membership structure.
The GP model, like other mixed models used in the value-added context
\citep{sanders97,rowan02,mc03,mc04,mc05,lock07}, uses a longitudinal database of student scores and models the scores with random teacher intercepts. Under this scenario, the empirical best linear unbiased predictors (EBLUPs) for random teacher intercepts are the
teacher VAM scores.
In this paper we use the term ``teacher effect'' to represent the VAM score of a teacher but note that,
as observed by \citet{lock07},
these
teacher effects measure ``unexplained heterogeneity at the classroom level,''
and not necessarily the causal effect of the teacher.

The GP model is distinguished from others in the VAM literature by how it attributes a student's performance
to current and prior teachers.
If the effects of good teaching persist, one would expect that students of a good teacher
in year 1 would do well on the test in year 1 and would continue to do well on the tests in future years.
The Educational Value-Added Assessment System (EVAAS) model \citep{sanders97}, a complete persistence model,
assumes that the effect of
a teacher persists undiminished over all subsequent years of his or her students' achievement.
This complete persistence assumption, also proposed by \citet{raudenbush02},
implies that each teacher has one VAM score: the effect of a teacher in year $g$ on his or her students'
test scores is the same for their tests in each of years $g,\ldots,T$.
The complete persistence assumption simplifies the covariance structure, and the EVAAS model is
implemented in SAS software \citep{wright}.
A model proposed by
\citet{mc04} allows the effect of a teacher on students' scores to decay in
future years, though the effects are otherwise perfectly correlated. \citet{lock07} refer to this structure as variable persistence (VP).
In the VP model, each teacher has one estimated effect, but the impact on students' future year
scores is reduced by a multiplicative factor in each year. The multipliers, called persistence parameters, are estimated from the data.

The GP model allows a much more general
structure for the effects of a current teacher on future test scores.
In the GP model, a teacher in year $g$ has a different effect on his or her students' scores in each year from $t = g,\ldots,T$,
and the $(T-g+1)$ effects of that teacher have an unstructured covariance matrix to allow the effects to be correlated.
The EVAAS model is a special case of the GP model in which the current and future effects of a teacher are assumed
to be identical.
The general correlation structure in the GP model allows much more detailed exploration of the patterns
of teacher effects, but greatly complicates the problem of computing estimates.

\citet{hill98} estimate a class of multiple membership models
using an iterative generalized least squares algorithm, and \citet{browne01}
employ Monte Carlo Markov chain techniques. Likewise, \citet{mariano10} use Bayesian methods to estimate the parameters for
the GP model using data from a large urban school district.
A Bayesian approach to computations requires a choice of prior distribution for the covariance parameters.
As investigated in their paper, different priors often result in different estimates of model parameters and teacher effects. A maximum likelihood (ML) approach avoids the need for priors, although ML estimation of even the simpler VP model 
has been ``practically infeasible for all but small data sets'' \citep{lock07} up to this point.
In this paper we use the sparseness of the covariance and design matrices to develop an efficient EM algorithm for calculating ML
estimates of parameters in
the GP and VP models.
We implement the method in the user-friendly GPvam package
\citep{gpvam} in R statistical software \citep{R}.
This development makes the GP and VP models more accessible for use in practice,
and provides an alternative to the Bayesian calculations implemented by \citet{mariano10}.

While the GP model was developed for educational applications, the model and the computational methods in this paper apply in many other settings as well. For example, \citet{ash} note the similarity between the problems of evaluating teacher performance on
the basis of student outcomes, and evaluating hospital and physician performance on the basis of patient outcomes.
The multiple membership structure also arises in social network data \citep{airoldi}.
In another example, \citet{browne01} and \citet{goldstein} describe a multiple membership model used to study Belgian household migration with complete persistence, measuring the propensity of individuals to change household membership. The GP model is a good candidate for the Belgian household data since the similarity of former roommates may decrease over time.
\citet{browne01} also describe an application in which a hospital patient is cared for by different nurses
and the contribution of each nurse to the patient's progress is estimated.

The paper is organized as follows.
Section \ref{sec:gp} reviews the models studied and lays out the foundation for ML estimation. Section \ref{sec:em} presents the EM algorithm for the estimation of the model. The details of the
implementation of the model in R appear in Section \ref{sec:implement}.
The computational methods are applied to a
data set from a large urban school district in Section \ref{sec:data} to demonstrate the capabilities of the estimation procedure and software.

\section{Model Specification}\label{sec:gp}

The GP model \citep{mariano10} and other structures considered in this paper model responses of the lower-level units as follows:
\begin{equation}\label{eq:model}
   y_{ig}=\bsx_{ig}^{\prime}\bsbeta+\bss_{ig}^{\prime}\bse+\epsilon_{ig}
\end{equation}
where $y_{ig}$ is a response for unit $i$ at time $g$ for $i=1,\ldots,n$, and $g\in A_i$; $A_i$, a subset of $\{1,\ldots,T\}$, is the set of times for which unit $i$ is observed. The vector of all responses is $\bds{y}=(\bds{y}_1^{\prime},\ldots,\bds{y}_{n}^{\prime})^{\prime}$, where $\bds{y}_i=(y_{ig}, g\in A_i)$ is the vector of responses for unit $i$.
The matrix $\bsX$, with rows $\bds{x}_{ig}^{\prime}$ for $g\in A_i$ and $i=1,\ldots,n$, is the design matrix of covariates for the fixed effects parameter vector $\bsbeta$.

The random effects vector $\bse\sim N(0,\bsG)$ contains random intercepts for the higher-level units (and also for the lower-level units
if desired). Each measurement on a lower-level unit is associated with multiple higher-level units as specified by the design matrix
 $\bsS$ which has rows $\bds{s}_{ig}^{\prime}$ for $g\in A_i$ and $i=1,\ldots,n$.
The multiple membership structure arises because rows of the $\bsS$ matrix may contain multiple nonzero values.
The vector of error terms for lower-level unit $i$, $\bds{\epsilon}_{i}=\left\{\epsilon_{ig}, {g\in A_i} \right\}$, is assumed to be
normally distributed with mean $\bds{0}$ and covariance matrix $\bsR_i$. The  lower-level units are assumed
to be independent conditionally on the random intercepts contained in $\bse$, so
$\bds{\epsilon} = (\bds{\epsilon}_{1}^{\prime},\ldots,\bds{\epsilon}_{n}^{\prime})^{\prime} \sim N(\bds{0},\bsR)$ and $\bsR$ is block diagonal with blocks $\bsR_1,\ldots,\bsR_n$. The error terms $\bds{\epsilon}$ are also assumed to be independent of the
effects in  $\bse$. 

The observations taken together thus have the general form of a linear mixed model:
\begin{equation}\label{eq:LMM}
\bsy=\bsX\bsbeta+\bsS\bse+\bds{\epsilon},
\end{equation}
with $\Cov(\bsy) = \bsV = \bsS \bsG \bsS^{\prime} + \bsR$.
The log-likelihood based on the observed data $\bsy$ from model (\ref{eq:LMM}) is
\begin{align}
l(\moparm;\bsy)={}&-\frac{N}{2}\log(2\pi)-\frac{1}{2}\log\left|\bds{V}\right|\nonumber\\
&-\frac{1}{2}\left(\bsy-\bsX\bsbeta\right)^{\prime}\bds{V}^{-1}\left(\bsy-\bsX\bsbeta\right),\label{eq:fulldatalikelihood}
\end{align}
where $N=\sum_i|A_i|$ is the number of observed responses and $\moparm$ is a vector of the unique model parameters from $\bds{\beta}$, $\bds{G}$, and $\bds{R}$, together with the free persistence parameters in the VP model.
We assume throughout that the sufficient conditions for consistency and asymptotic normality of the ML estimates given by
\citet{broatch10} are met.
In addition to usual regularity conditions, \citet{broatch10} assume that $T$ is bounded and that the number of lower-level units associated
with each higher-level unit is bounded, i.e., that the sum of each column of $\bsS$ is bounded.
The model is also assumed to be identifiable, with $\bsX$ of full column rank. For the models with fixed $\bsS$, let $\psi_1,\ldots,\psi_q$ be the unique covariance parameters and write
$\bsV = \sum_{j=1}^q \psi_j \bsSigma_j$.
These covariance parameters are identifiable when the matrices $\bsSigma_j$ are linearly independent for $j=1,\ldots,q$.
For the VP model, identifiability of the free persistence parameters must also be checked because $\bsS_*$ depends on them.
In practical terms, identifiability depends on the observation pattern and on sufficient mixing in the population, so that students in a class progress to a variety of different teachers as they continue through school.
\citet{briggs11}, fitting a VP VAM with schools as higher-level units instead
of teachers, find that not all parameters are identifiable because 
most students in their data set move from one grade to the next as a cohort within the same school with insufficient mixing.

Note that the model formulation allows lower-level units to be missing observations for some times.
In this paper we assume that observations are missing at random and that the parameters governing the outcome process are distinct from those characterizing the missingness process, yielding a valid likelihood-based analysis under the specified model \citep{little87}. 
\citet{mc10} and \citet{karl} propose joint models for  test score data and missingness indicators to accommodate data with informative missingness, but we do not consider such models here.

We handle the missing teacher links resulting from missing student observations by assuming that the student was taught by an average teacher in that year. For example, when modeling scores from grades 1, 2, and 3, if a student enters the school at grade 2, we do not link that student's second and third grade scores to any of the first grade teachers. This approach was also used by \citet{lock07}.

To make these ideas concrete, in the remainder of this section we present specific models  considered in the educational setting, in which the lower-level units are students,
the higher-level units are teachers, and the measurement $y_{ig}$ is a test score of student $i$ in year $g$.
Rather than introducing new notation for each model, we recast the model terms to match the notation of Equations (\ref{eq:model})
and  (\ref{eq:LMM}) so that the definitions of $\bse,\bsG,\bsR$ and $\bsS$ depend on the chosen model. This streamlines the discussion of the estimation of the parameters.

\subsection{Generalized Persistence Model}\label{ssec:gp}	

The GP VAM \citep{mariano10} models student scores using information about the history of observations on each student and each student's teacher-history.  It estimates the effect of teachers on students in the year that they teach them, their lasting effect on the next year's score, and so on. Following the notation of \citet{mariano10}, let $\theta_{g[jt]}$ represent the effect for the $j$-th grade-$g$ teacher on a student's grade $t$ score, for $t\geq g$. A grade $g$ teacher has $K_g=T-g+1$ effects, one each for grades $g,\ldots,T$. Thus $\bds{\theta}_{g[j\cdot]}$ gives the vector of current and future year effects of the $j$-th grade $g$ teacher. The vector $\bse$ concatenates the $\bds{\theta}_{g[j\cdot]}$ effects for all grades and teachers. The model is able to distinguish between the persistence effect of former teachers and the current effect of the present teacher because the students are not nested in teachers.

We structure $\bse$ so that $\bsG$ will be block diagonal: if
\begin{equation}\label{eq:eta1}
\bds{\eta}=(\bds{\theta}^{\prime}_{1[1\cdot]},\ldots,\bds{\theta}^{\prime}_{1[m_1\cdot]},\bds{\theta}^{\prime}_{2[1\cdot]},\ldots,\bds{\theta}^{\prime}_{2[m_2\cdot]},\ldots,{\theta_{T[1\cdot]}},\ldots, {\theta_{T[m_T\cdot]}})^{\prime}
\end{equation}
then $\bds{\eta} \sim N(\boldsymbol{0},\boldsymbol{G})$, where
\begin{equation}\label{eq:G}
\bds{G}=\te{blockdiag}\left(\bsGamma_1,\ldots,\bsGamma_1,\ldots,\Gamma_T,\ldots,\Gamma_T\right).
\end{equation}
With $m_g$ teachers in year $g$, there are $m_g$ copies each of $\bsGamma_g$ and each $\bsGamma_g$ is unstructured. The matrix $\bsGamma_g$ is square with $K_g$ rows and gives the covariance of current and future year effects for teachers of grade $g$. The vector $\bss_{ig}$  contains $1$'s in entries corresponding to teachers who could affect response $g$ of student $i$.
Thus, for a measurement $y_{i2}$, where student $i$ had teacher 5 at time 1 and teacher 12 at time 2, $\bss_{i2}$ contains a 1
corresponding to the position of $\bds{\theta}_{1[5,2]}$ to include the lagged-year effect of teacher 5, and it contains a 1 corresponding
to the position of $\bds{\theta}_{2[12,2]}$ to include the current-year effect of teacher 12.
If, on average, teachers have more effect on current-year student scores than on subsequent scores of their students, we expect
the first diagonal element of $\bsGamma_g$ to be larger than the other diagonal elements, reflecting the larger variability
of current-year teacher effects.

The intra-student correlation is modeled in unstructured blocks of the conditional covariance matrix $\bsR$.
After ordering the data by student and then by year, the error terms $\bds{\epsilon} =( \bds{\epsilon}_1', \ldots, \bds{\epsilon}_n')'$ are distributed as $\bds{\epsilon} \sim N(\bds{0},\bds{R})$ where $\bds{R}$ is a block diagonal matrix with blocks
\begin{align}\label{eq:R}
\bsR_i=\begin{pmatrix}
\sigma_{11}&\cdots&\sigma_{T1}\\
\vdots&\ddots&\vdots\\
\sigma_{T1}&\cdots&\sigma_{TT}
\end{pmatrix}.
\end{align}
If student $i$ is missing an observation, then $\bsR_i$ omits the corresponding row and column corresponding to the year in which the observation is missing. $\bsR_i$ depends on $i$ through the observed-year pattern $A_i$. We refer to this model as GP.R, indicating that the intra-student correlation is modeled in the $\bsR$ matrix.

An advantage of this model is that the responses in different years can use different scales---the scaling
is picked up in the covariance matrices $\bsG$ and $\bsR$. The model has great flexibility for the relation between
current- and future-year teacher effects, and for the within-student correlation. Note that this formulation assumes that the sets of teachers in different grades are distinct (or that if someone teaches
in both grades 3 and 4, their effects on the grade-3 and grade-4 students are independent). The model can be modified
to allow additional dependence for persons who teach multiple grades, but for simplicity here we consider the case with distinct teachers.

\subsection{GP Model with a Single Future Year Effect}\label{sec:rGP}
Some processes, including the educational data analyzed by \citet{mariano10}, produce strongly correlated future year effects. \citet{mariano10} note in their application that, within each grade, the future year effects are strongly correlated with each other, but only moderately correlated to the current year effect. Following their idea of averaging the future year effects of each teacher after fitting the full model, we fit a reduced model that allocates a single future year effect to each teacher. This combines aspects of the GP model and the complete persistence model.
This reduction requires that the scale of measurement be the same for each year of the study.
We refer to the reduced model as rGP.R. We use $\theta_{t[j1]}$ and $\theta_{t[j2]}$ to represent the current and future year effects, respectively, of the $j$-th grade-$t$ teacher in rGP.R.
Then $\bss_{ig}$ contains a 1 in the entry corresponding to the teacher in year $g$, $\theta_{g[j1]}$, and also contains a 1 in
each entry corresponding to the future effects of the student's teachers in years $1,\ldots,g-1$.
An alternative model would be to impose an autoregressive structure on the $\bsGamma_i$ as in
\citet{paddock11}.

\subsection{GP Model with Random Student Effects}\label{sec:alt}
When scores from each year are measured on the same scale, an alternative model specification is available. Using a variable persistence structure for teacher effects, \citet{mc10} modeled the intra-student correlation by using a random intercept for each student. We implement this alternative structure here, except we use the generalized persistence structure for teacher effects. We refer to this model as GP.G:
\begin{equation}\label{eq:altmodel}
   y_{ig}=\bsx_{ig}^{\prime}\bsbeta+\bss_{ig}^{\prime}\bse_*+\delta_i+\epsilon_{ig}
\end{equation}
The terms in Equation (\ref{eq:altmodel}) are defined the same as they were in Equation (\ref{eq:model}), with the exception of $\epsilon_{ig}$ and the new term $\delta_i$. For this subsection, we use $\bse_*$ to denote the vector of teacher effects. Instead of modeling $\bds{\epsilon}_i$ with an unstructured covariance matrix, GP.G includes a separate error variance in each year $\epsilon_{ig}\sim N(0,\sigma^2_g)$. As a result, $\bsR$ is diagonal with entries from the set $\left\{\sigma^2_1,\ldots,\sigma^2_T\right\}$, corresponding to the year of the observation. We likewise offer new definitions for $\bsG, \bsS$ and $\bse$ for GP.G.

The $\delta_i$ are random student intercepts, distributed as $\delta_i\sim N_1(0,\Gamma_{stu})$, with $\te{cov}(\epsilon_{ig},\delta_i) = 0$. We may express GP.G in the form of Equation (\ref{eq:model}) by including the $\delta_i$ in the random effects vector $\bse$,
\begin{equation}\label{eq:eta2}
\bds{\eta}=(\delta_1,\ldots,\delta_n,\bse_*^{\prime})^{\prime}.
\end{equation}
The vector $\bds{\eta}$ is then distributed as $\bds{\eta} \sim N(\boldsymbol{0},\boldsymbol{G})$ where
\begin{equation}\label{eq:G2}
\bds{G}=\te{blockdiag}\left(\Gamma_{stu}\te{\textbf{I}}_n,\bsGamma_1,\ldots,\bsGamma_1,\ldots,\bsGamma_T,\ldots,\bsGamma_T\right),
\end{equation}
with $m_g$ copies each of $\bsGamma_g$, where each $\bsGamma_g$ is unstructured. To accommodate the new $\bse$, the design matrix $\bsS$ is composed of the blocks $[\bsS_1|\bsS_2]$, where $\bsS_1$ is the design matrix for the student effects and $\bsS_2$ is the design matrix for the teacher effects.

The same model could be fit without student random intercepts by using a block-diagonal $\bsR$ with student blocks
\[
\Gamma_{stu}\bds{1}\bds{1}^{\prime}
+\te{diag}(\sigma_g^2:g\in A_i).
\]
These blocks are compound symmetric when the year-specific residual variances are equal. However, the student-intercept formulation is useful when exploring sensitivity to the presence of potentially nonignorable missing data \citep{mc10,karl} or when the random student intercepts are of interest.
The GP.G formulation is also more easily extended to allow a random growth model where each student has his or her own
slope and intercept.

\subsection{Complete and Variable Persistence Models}\label{ssec:CP}
Instead of modeling a separate effect in years $g,\ldots, T$ for each grade-$g$ teacher, the variable persistence (VP) VAM models a single effect for each teacher. Let $\theta_{t[j]}$ denote the effect of the $j$-th grade-$t$ teacher. The persistent effect of the $j$-th grade-$t$ teacher on grade-$g$ scores is modeled as a multiple of that teacher's effect, $\alpha_{gt}\theta_{t[j]}$. \citet{lock07} refer to the $\alpha_{gt}$ for $g=1,\ldots,T$ and  $t=1,\ldots,g$ as persistence parameters. The persistence parameters for the current year are fixed at one, $\alpha_{gt}=1$ for $t=g$, while the others are estimated.
The complete and zero persistence VAMs are two special cases of the VP model, with fixed persistence parameters $\alpha_{gt}=1$ and  $\alpha_{gt}=0$ (for $t\neq g$), respectively. The $\bsR$ matrix of VP is the same as the one defined for GP.R.
SAS does not provide the ability to estimate the VP model:  \citet{lock07} note that there are no available scalable implementations of the VP model. However, we will show that the EM algorithm can provide a scalable routine for the VP model.

The random teacher effects for the VP model are concatenated
\begin{equation*}
\bse=\left(\theta_{1[1]},\ldots,\theta_{1[m_1]},\theta_{2[1]},\ldots,\theta_{2[m_2]},\ldots,\theta_{T[1]},\ldots,\theta_{T[m_T]}\right).
\end{equation*}
and distributed as $N(0,\bsG)$.
Since there is only one effect modeled for each teacher, $\bsG$ is diagonal with $m_g$ copies of $\Gamma_g$ for $g=1,\ldots,T$,
\begin{equation*}
\bsG=\te{diag}\left(\Gamma_1,\ldots,\Gamma_1,\Gamma_2,\ldots\Gamma_2,\ldots,\Gamma_T,\ldots,\Gamma_T\right),
\end{equation*}

In the VP and CP models, the $\alpha_{tg}$'s modify the covariance structure,
since the effect of teacher $j$ in year $g$, $\theta_{g[j]}$, appears in the model for that teacher's students in all subsequent years. To match the structure of the VP model to that of the GP model, let $\bse_*=\bsA\bse$ where $\bsA$ has $\sum_g m_g$ columns, one for each teacher. $\bsA=\te{blockdiag}(\bsA_1,\ldots,\bsA_T)$, where $\bsA_g=\bds{I}_{m_g}\otimes\bds{a}_g$, $\bds{a}_g=(\alpha_{tg})_{t=g}^{T}$ is a column vector with $\alpha_{gg}=1$, for $g=1,\ldots,T$, with $\otimes$ representing the Kronecker product. Using the same definition of $\bsS$ as in Section \ref{ssec:gp}, the VP model may be expressed as

\begin{equation}\label{eq:VPmodel}
\bsy=\bsX\bsbeta+\bsS\bse_*+\bds{\epsilon},
\end{equation}
where $\bse_*\sim N(\bds{0},\bsG_*)$, with $\bsG_*=\bsA\bsG\bsA^{\prime}$. The error terms are distributed as $\bds{\epsilon}\sim N(\bds{0},\bsR)$, $\te{cov}(\bse,\bds{\epsilon})=0$, where $\bsR$ is defined in Section \ref{ssec:gp}. \citet{briggs11} discuss alternative formulations of the VP model that can be used if there are concerns about identifiability.

\section{Computing Maximum Likelihood Estimates}\label{sec:em}

\label{sec:em1}

The degree of computational difficulty associated with estimating the parameters of Model (\ref{eq:LMM}) depends largely on the structure of the random effects, manifested through the pattern of nonzero entries in $\bsS$. In applications where the random effects are nested within subjects, the resulting $\bds{V}$ matrix is block diagonal, and the log-likelihood in (\ref{eq:fulldatalikelihood}) may be factored over the subjects. However, for non-nested models, $\bsV$ has no patterned structure, and its dimension is equal to the number of observations in the data set. As a result, a direct maximization of the likelihood function is highly inefficient or infeasible for large data sets. \citet{wolf94} develop a dimensionality-reduction technique---used with a Newton-Raphson (NR) routine in SAS$^\circledR$ software \citep{sas}---that requires the manipulation of a square matrix with dimension depending on the number of levels of fixed and random effects, rather than the number of observations. Either the method of \citet{wolf94} or some other form of dimensionality reduction  is necessary for scalable estimation of Model (\ref{eq:LMM}) when the random effects are not nested.

Even after a dimensionality reduction for $\bsV$, the matrices $\bsR$ and $\bsS$ grow with the size of the data set. For example, SAS PROC GLIMMIX allows users to specify a custom $\bsS$ matrix via the multimember option of its EFFECT statement. However, the procedure does not currently take into account the sparse structure of the design and covariance matrices, and does not scale well to large data sets. SAS PROC HPMIXED does take sparseness into account and can be used to estimate a
variation of the complete persistence model. However, HPMIXED is tailored to a specific model and has limited choice of covariance structures.
\citet{broatch10} show how to use SAS software to estimate parameters in a multiresponse VAM
by specifying a user-defined covariance matrix, but this method also does not work well with large data sets.

Model (\ref{eq:LMM}) requires a positive definite $\bsG$ matrix.  A third major difficulty associated with the estimation of linear mixed models arises when random effects are highly correlated, producing a nearly singular $\bsG$ matrix. The Newton-Raphson routines are prone to failure in these settings, frequently producing non-positive definite estimates for $\bsG$ \citep{demidenko}. One possible solution to this issue is by parameterizing the model according to the Cholesky root of $\bsG$. SAS offers functionality for such a parametrization, but it is only compatible with banded-unstructured covariance matrices \citep{sas}.

The EM algorithm presented below overcomes these challenges by using a matrix of reduced dimension from that of $\bsV$,
utilizing the sparseness of $\bsS, \bsG$, and $\bsR$, and achieving stability when the random effects covariance matrix is nearly singular.

\subsection{The EM Algorithm}\label{sec:em2}

The EM algorithm \citep{dempster77,embook} provides a broad framework for maximum likelihood estimation in the presence of missing data. It was one of the first methods used to estimate linear mixed models by treating latent random effects as missing data \citep{laird82}. Its use for estimation of mixed models has lagged behind the popularity of the often-faster NR algorithms. The EM algorithm has a linear rate of convergence (which depends on the number and structure of the random effects) near a local maximum \citep{dempster77}, whereas the NR algorithms provide a quadratic rate of convergence.

Nevertheless, an advantage of the EM algorithm is that no restrictions need to be placed on the $\bsG$ matrix to ensure that it is positive definite, as shown in the Appendix. For some models, the usual advantages of NR over EM \citep{lindstrom88} are negated by the presence of highly correlated random effects. Furthermore, the EM algorithm naturally depends on the manipulation of matrices of dimension equal to the number of random effects rather than the number of observations so that additional dimensionality reduction techniques are not necessary.  When taking advantage of sparse matrix computations, the EM algorithm can provide a viable method for estimating non-nested mixed models, especially those with highly correlated random effects.

We will refer to $f(\bsy; \moparm)$ as the observed data density function and \newline $f(\bsy, \bse; \moparm) = f(\bsy|\bse; \moparm) f(\bse; \moparm)$ as the complete data density function, where
\small
\begin{align*}
f(\bsy|\bse; \moparm) &\propto \left|\bsR\right|^{-1/2}\exp\left\{-\frac{1}{2}\left(\bsy-\bsX\bsbeta -\bsS\bse\right)^{\prime}\bsR^{-1}\left(\bsy-\bsX\bsbeta -\bsS\bse\right)\right\}\\
f(\bse; \moparm) &\propto \left|\bsG\right|^{-1/2}\exp\left\{-\frac{1}{2}\bse^{\prime}\bsG^{-1}\bse\right\}
\end{align*}
\normalsize
Given initial values for the parameters and the random effects, the EM algorithm alternates between an expectation (E) step and a maximization (M) step. At iteration (k + 1), the E step calculates the conditional expectation of the complete data log-likelihood, given the observed data, $\bsy$, and parameter estimates obtained in the k-th step, $\moparm^{(k)}$. That is, the E step computes
\begin{equation*}
Q(\moparm; \moparm^{(k)}) = \int{\left\{\log f\left(\bsy|\bse; \moparm\right) + \log f\left(\bse; \moparm\right)\right\} f(\bse|\bsy; \moparm^{(k)})} \ud\bse.
\end{equation*}
The M step then maximizes $Q(\moparm; \moparm^{(k)})$ with respect to $\moparm$, resulting in an updated parameter vector $\moparm^{(k + 1)}$. Any interior M-step maximizer satisfies
\begin{equation}
\int \frac{\partial}{\partial\moparm} \left\{ \log f(\bsy|\bse; \moparm) + \log f(\bse; \moparm)\right\} f(\bse|\bsy; \moparm^{(k)} )\ud\bse\Big|_{\moparm=\moparm^{(k+1)}} = \bds{0}\label{eq:emscore1},
\end{equation}
provided that differentiation and integration are interchangeable, which is valid because the complete data likelihood $f(\bsy,\bse; \moparm)$ is a member of the exponential family \citep{lehmann}. The left side of Equation (\ref{eq:emscore1}) is the gradient of $Q$ with its second argument held fixed. The observed data score instead satisfies the identity
\[
S(\moparm;\bsy)
=\left.\frac{\partial Q(\bds{\xi};\moparm)}{\partial\bds{\xi}}
\right|_{\bds{\xi}=\moparm}
=\te{E}_{\moparm}\!\left[
\frac{\partial\log f(\bsy,\bse;\moparm)}{\partial\moparm}
\,\middle|\,\bsy\right]
\]
\citep{louis}. The M-step gradient vanishes at an interior maximizer, but the observed data score need not vanish until convergence.

In Sections~\ref{ssec:mg} to \ref{ssec:MVP} we derive the M step for each of the models developed in Section \ref{sec:gp}. In these derivations, $U$ denotes a derivative of $Q$ with the conditional moments held fixed at $\moparm^{(k)}$; we refer to $U$ as a score function in this sense. Coupled parameter blocks may be updated by conditional maximization, holding the other blocks fixed and ensuring that each update increases or preserves $Q$.  The E step described in Section~\ref{sec:estep} is the same for all of the models discussed, using the appropriate definitions of $\bse$, $\bsG$, $\bsS$, and $\bsR$.

\subsection{M-Step for GP.R and rGP.R} \label{ssec:mg}
The M-step updates appearing in this section apply to both the generalized persistence model GP.R and its reduced version rGP.R. The only differences that must be kept in mind are the definitions of $\bse$ and $\bsGamma_g$. Using the definition of $\bsG$ in Equation (\ref{eq:G}), which applies to both GP.R and rGP.R, we may write the density of $\bse$ as
\begin{eqnarray*}
f(\bse;\moparm)&\propto&\det(\bds{G})^{-1/2}\exp\left(-\frac{\bse^{\prime}\bds{G}^{-1}\bse}{2}\right)\\
&=& \left[\prod_{g=1}^T\det(\bds{\Gamma_{\te{g}}})^{-m_g/2}\right]\exp\left(-\sum_{g=1}^T\sum_{j=1}^{m_g}\frac{\bstheta_{g[j\cdot]}^{\prime}\bds{\Gamma}_{g}^{-1}\bstheta_{g[j\cdot]}}{2}\right)
\end{eqnarray*}
We use \citet*{cook} and \citet*{har} for matrix differentiation, and note that each $\bsGamma_{g}$ is symmetric. Referring to Equation (\ref{eq:emscore1}), the score vector with respect to $\bsGamma_{g}$ is
\begin{eqnarray*}
U(\bds{\Gamma_{g}})&=&\int \der{\bds{\Gamma_{g}}}\log\left[\det(\bds{G})^{-1/2}\exp\left(-\frac{\bse^{\prime}\bds{G}^{-1}\bse}{2}\right)\right]f(\bse|\bsy;\moparm^{(k)})\ud \bse\\
&=&-\frac{1}{2}\int\der{\bds{\Gamma_{g}}}\left\{m_g\log\left[\det(\bds{\Gamma_{g}})\right]+\sum_{j=1}^{m_g}\bstheta_{g[j\cdot]}^{\prime}\bds{\Gamma}_{g}^{-1}\bstheta_{g[j\cdot]}\right\}f(\bse|\bsy;\moparm^{(k)})\ud \bse\\
&=&\textrm{matrix with components }
\left\{ \begin{array}{ll}
d_{ij} & \textrm{if $i=j$}\\
2d_{ij} & \textrm{if $i\neq j$}
\end{array} \right.
\end{eqnarray*}
where $d_{ij}$ is the $ij$-th component of the matrix
\begin{equation*}
\bds{D}=-\frac{1}{2}\left\{m_g\bds{\Gamma_{g}}^{-1}-\bds{\Gamma_{g}}^{-1}\left(\sum_{j=1}^{m_g}\te{E}\left[\bstheta_{g[j\cdot]}\bstheta_{g[j\cdot]}^{\prime}|\bsy;\moparm^{(k)}\right]\right)\bds{\Gamma_{g}}^{-1}\right\}
\end{equation*}
Let
\begin{align}
\widetilde{\bse}&=\te{E}[\bse|\bsy;\moparm^{(k)}]\label{eq:etatilde}\\
\widetilde{\bsv}&=\te{var}[\bse|\bsy;\moparm^{(k)}]\label{eq:vtilde}
\end{align}
represent the conditional expectation and variance, respectively, of $\bse$. These quantities are calculated in the E-step and remain fixed during the M-step. Likewise, let the sub-vector of $\widetilde{\bse}$ corresponding to $\te{E}[\bstheta_{g[j\cdot]}|\bsy;\moparm^{(k)}]$ be denoted $\widetilde{\bstheta}_{g[j\cdot]}$, and the block of the matrix $\widetilde{\bsv}$ corresponding to $\te{var}[\bstheta_{g[j\cdot]}|\bsy;\moparm^{(k)}]$ be denoted $\widetilde{\bsv}_{g[j\cdot]}$.
Now, since $\widetilde{\bsv}=\te{E}[\bse\bse^{\prime}|\bsy;\moparm^{(k)}]-\widetilde{\bse}\widetilde{\bse}^{\prime}$, setting $U(\bsGamma_{g})=\bds{0}$ implies
\begin{eqnarray*}
m_g\bds{\Gamma_{g}}^{-1}&=&\bds{\Gamma_{g}}^{-1}\sum_{j=1}^{m_g}\left(\widetilde{\bsv}_{g[j\cdot]}+\widetilde{\bstheta}_{g[j\cdot]}\widetilde{\bstheta}_{g[j\cdot]}^{\prime}\right)\bds{\Gamma_{g}}^{-1}
\end{eqnarray*}
Thus the M-step update for $\bds{\Gamma_{g}}$ is
\begin{equation}\label{eq:gammag}
\widehat{\bds{\Gamma}}_{g}=\frac{1}{m_g}\sum_{j=1}^{m_g}\left(\widetilde{\bsv}_{g[j\cdot]}+\widetilde{\bstheta}_{g[j\cdot]}\widetilde{\bstheta}_{g[j\cdot]}^{\prime}\right)
\end{equation}
Equation (\ref{eq:gammag}) calculates an average of the blocks of $\widetilde{\bsv}+\widetilde{\bse}\widetilde{\bse}^{\prime}$ that correspond to teachers who taught in year $g$.

The M-step update for $\bsbeta$ is the value that solves $U(\bsbeta)=0$, where
\begin{align*}
U\left(\bsbeta\right)&=\int\der{\bsbeta}\left[-\frac{1}{2}\left(\bsy-\bsX\bsbeta-\bsS\bse\right)^{\prime}\bsR^{-1}\left(\bsy-\bsX\bsbeta-\bsS\bse\right)\right]f(\bse|\bsy;\moparm^{(k)})\ud \bse\\
&=\bsX^{\prime}\bsR^{-1}\left(\bsy-\bsX\bsbeta-\bsS\widetilde{\bse}\right),
\end{align*}
namely,
\begin{align}\label{eq:beta}
\widehat{\bsbeta}=\left(\bsX^{\prime}\bsR^{-1}\bsX\right)^{-1}\bsX^{\prime}\bsR^{-1}\left(\bsy-\bsS\widetilde{\bse}\right)
\end{align}

The calculation of the M-step update for $\bsR$ from Equation (\ref{eq:R}) is complicated by the fact that the structure of $\bsR$ changes in the presence of unbalanced data. The M-step update for the component $\sigma_{kl}$ of $\bsR$ is the value that solves $U(\sigma_{kl})=0$, where
\begin{align*}
U\left(\sigma_{kl}\right)&=\int\der{\sigma_{kl}}\left[\vphantom{\frac{1}{2}}\log\left(\left|\bsR\right|^{-1/2}\right)\right.\\
{}&\left.-\frac{1}{2}\left(\bsy-\bsX\bsbeta-\bsS\bse\right)^{\prime}\bsR^{-1}\left(\bsy-\bsX\bsbeta-\bsS\bse\right)\right]f(\bse|\bsy;\moparm^{(k)})\ud \bse.
\end{align*}

If the observations are sorted by students and then by year, $\bsR$ is block-diagonal with block sizes depending on the number of observations on each student. For $T$ years, there are $2^T-1$ possible combinations of years in which a student may be observed, although not all of these patterns may appear in a given data set. To parameterize these combinations, we treat the ordered, binary observed-test-score (OTS) indicators for each student as a number in base-2. So in a study over three years, each student will have an OTS pattern from the first column of Table \ref{tab:pat}.

\begin{table}
\caption{Parameterizing the OTS patterns for example with 3 years}
\centering
\label{tab:pat}
\begin{tabular}{lr}
\toprule
OTS\\ indicators& Pattern\\
\midrule
001&1\\
010&2\\
011&3\\
100&4\\
101&5\\
110&6\\
111&7\\
\bottomrule
\end{tabular}
\end{table}

For example, a student with observations in each year has pattern 7, with the corresponding block of $\bsR$ given by
\begin{displaymath}
\begin{pmatrix}
\sigma_{11}&\sigma_{21}&\sigma_{31}\\
\sigma_{21}&\sigma_{22}&\sigma_{32}\\
\sigma_{31}&\sigma_{32}&\sigma_{33}
\end{pmatrix}.
\end{displaymath}
The matrices corresponding to the other patterns are subsets of this matrix, using the rows and columns suggested by the OTS indicator. A student who is missing an observation in year 2 has pattern 5 and corresponding error covariance matrix
\begin{displaymath}
\begin{pmatrix}
\sigma_{11}&\sigma_{31}\\
\sigma_{31}&\sigma_{33}
\end{pmatrix}.
\end{displaymath}

Let $p$ denote the OTS pattern, $n_p$ be the number of students with that pattern, and $\bsR_{(p)}$ represent the covariance matrix corresponding to the $p$-th pattern. In addition, let $P_{kl}$ denote the set of patterns $p$ whose covariance matrix $\bsR_{(p)}$ contains $\sigma_{kl}$. Furthermore, let $b(p)$ denote the $b$-th student with pattern $p$. We may write
\begin{displaymath}
\left|\bsR\right|=\prod_p \left|\bsR_{(p)}\right|^{n_p}.
\end{displaymath}
Thus the score function may be expressed as
\begin{align*}
U\left(\sigma_{kl}\right)=&-\frac{1}{2}\int\der{\sigma_{kl}}\left\{\sum_p n_p\log\left|\bsR_{(p)}\right|+\sum_p\sum_b \left[ \vphantom{\bsR_{(p)}^{-1}} \right.\right.\\
&\left. \left. \left(\bsy_{b(p)}-\bsX_{b(p)}\bsbeta-\bsS_{b(p)}\bse\right)^{\prime}\bsR_{(p)}^{-1}\left(\bsy_{b(p)}-\bsX_{b(p)}\bsbeta-\bsS_{b(p)}\bse\right)\right]\vphantom{\sum_p}\right\}\\
&\times f(\bse|\bsy;\moparm^{(k)})\ud \bse
\end{align*}
where $\bsy_{b(p)}$ is the vector of observations from student $b(p)$, with corresponding design matrices for fixed and random effects $\bsX_{b(p)}$ and $\bsS_{b(p)}$. The derivative will be 0 for all terms that do not contain the parameter $\sigma_{kl}.$ This includes observations on students who do not have observations in both years $k$ and $l$. Then, taking the derivative and letting $1_{\{\mathcal{C}\}}$ be the indicator function that takes the value 1 if condition $\mathcal{C}$ is true and 0 otherwise,
\begin{align*}
U\left(\sigma_{kl}\right)=&{}-\left(1_{\{k\neq l\}}+\frac{1}{2}\times 1_{\{k=l\}}\right) \sum_{p\in P_{kl}}\left\{n_p \left(\bsR_{(p)}^{-1}\right)_{\left\{kl\right\}}-\vphantom{\int \sum_b\left[\bsR_{(p)}^{-1}\left(\bsy_{b(p)}-\bsX_{b(p)}\bsbeta-\bsS_{b(p)}\bse\right)\right.} \right.\\
&{}\int\sum_b \left[\bsR_{(p)}^{-1}\left(\bsy_{b(p)}-\bsX_{b(p)}\bsbeta-\bsS_{b(p)}\bse\right)\right.\\
&\left.\left.\times \left(\bsy_{b(p)}-\bsX_{b(p)}\bsbeta-\bsS_{b(p)}\bse\right)^{\prime}\bsR_{(p)}^{-1}\right]_{\left\{kl\right\}}f(\bse|\bsy;\moparm^{(k)})\ud \bse \right\}.
\end{align*}
The notation $\left\{kl\right\}$ indicates the matrix component corresponding to the position of the parameter $\sigma_{kl}$ in $R_{(p)}$. Again using the relationship $\widetilde{\bsv}=\te{E}[\bse\bse^{\prime}|\bsy;\moparm^{(k)}]-\widetilde{\bse}\widetilde{\bse}^{\prime}$,
\begin{align}
U\left(\sigma_{kl}\right)=&{}-\left(1_{\{k\neq l\}}+\frac{1}{2}\times 1_{\{k=l\}}\right)\sum_{p\in P_{kl}}\Bigg\{  n_p \bsR_{(p)}^{-1}\nonumber\\
&-\bsR_{(p)}^{-1}\sum_b\left[ \left(\bsy_{b(p)}-\bsX_{b(p)}\bsbeta\right)\left(\bsy_{b(p)}-\bsX_{b(p)}\bsbeta\right)^{\prime}\right.\nonumber\\
&- \left(\bsy_{b(p)}-\bsX_{b(p)}\bsbeta\right)\left(\bsS_{b(p)}\widetilde{\bse}\right)^{\prime}- \bsS_{b(p)}\widetilde{\bse}\left(\bsy_{b(p)}-\bsX_{b(p)}\bsbeta\right)^{\prime}\nonumber\\
&+\left.\bsS_{b(p)}\left(\widetilde{\bsv}+\widetilde{\bse}\widetilde{\bse}^{\prime}\right)\bsS_{b(p)}^{\prime}\vphantom{\left(\bsy_{b(p)}-\bsX_{b(p)}\bsbeta\right)^{\prime}}\right]\bsR_{(p)}^{-1}\Bigg\}_{\left\{kl\right\}}.\label{eq:mstepr}
\end{align}

If there were no missing observations then there would only be one OTS pattern and the calculation of the M-step update for $\bsR$ would have a solution that followed the same pattern as the M-step update for $\bsG$. However, the presence of unbalanced student profiles disrupts the structure of $\bsR$, and score functions must be calculated for each of the unique model parameters in $\bsR$. The closed form solution for $U(\sigma_{kl})=0$ depends on the number of years and on the OTS patterns that are present in the data set. One option is to use a Newton-Raphson routine to solve the score equations. We suggest such a method in Section \ref{sec:implement}.

\subsection{M-step for GP.G}
The M-step update for $\bsbeta$ in GP.G is the same as the update for GP.R appearing in Equation (\ref{eq:beta}), given the appropriate definition of $\bsR$. Likewise, the M-step updates for the $\bsGamma_g$ appearing in Equation (\ref{eq:gammag}) are unchanged.

The new work required for GP.G in Equation (\ref{eq:altmodel}) is the calculation of the M-step update for the student variance component $\Gamma_{stu}$ and the yearly error variances $\sigma^2_g$, for $g=1,\ldots,T$. The M-step update for $\Gamma_{stu}$ is derived in the same way as the update for $\bsGamma_g$, and is equal to the mean of the first $n$ diagonal elements of $\widetilde{\bsv}+\widetilde{\bse}\widetilde{\bse}^{\prime}$. For the purpose of calculating $\widehat{\sigma}_g^2$, let $B_g$ be the set of students that are observed in year $g$, and let $n_g=|B_g|$.

\small
\begin{align*}
U(\sigma^2_g)&=\int\der{\sigma^2_g}\left[\log\left(\prod_{j=1}^T\prod_{i\in B_j}\sigma^{-1}_j\exp\left[-\frac{\left(y_{ij}-\bsx^{\prime}_{ij}\bsbeta-\bss^{\prime}_{ij}\bse\right)^2}{2\sigma^2_j}\right]\right)\right]f(\bse|\bsy;\moparm^{(k)})\ud \bse\\
\end{align*}
\normalsize
Setting the score function equal to 0 and then using the fact that
\begin{displaymath}
\te{E}\left[{\bse^{\prime}\bss_{ig}}\bss^{\prime}_{ig}\bse|\bsy;\moparm^{(k)}\right]=\te{tr}\left({\bss_{ig}}\bss^{\prime}_{ig}\widetilde{\bsv}\right)+\widetilde{\bse}^{\prime}{\bss_{ig}}\bss^{\prime}_{ig}\widetilde{\bse}
\end{displaymath}
yields
\begin{align}
\widehat{\sigma}^2_g&=\frac{1}{n_g}\sum_{i\in B_g}\left\{\left(y_{ig}-\bsx^{\prime}_{ig}\bsbeta\right)\left(y_{ig}-\bsx^{\prime}_{ig}\bsbeta-2\bss^{\prime}_{ig}\widetilde{\bse}\right)+\bss^{\prime}_{ig}\left(\widetilde{\bsv}+\widetilde{\bse}\widetilde{\bse}^{\prime}\right){\bss_{ig}}\right\}\label{eq:trace}
\end{align}

\subsection{M-step for VP and CP Models}\label{ssec:MVP}
Although the covariance structure of the VP model in Equation (\ref{eq:VPmodel}) is different, the parameters may be estimated in much the
same way as for the GP model. The EM algorithm requires a positive definite covariance matrix for the random effects. Since $\bsG_*$ in Equation (\ref{eq:VPmodel}) is singular, we work instead with the diagonal $\bsG$ matrix defined in Section \ref{ssec:CP} and the associated vector $\bse$ of current year teacher effects.
This is done operationally by forming $\bsS_* = \bsS\bsA$, so that the ``design'' matrix $\bsS_*$ includes the parameters $\alpha_{gt}$, and
then iteratively updating  $\bsS_*$  as the parameter estimates are updated during the estimation procedure. This is merely an algebraic distinction, since $\bsS\bsA\bse=\bsS\bse_*=\bsS_*\bse$, where $\bse_*$ is the vector defined in Section \ref{ssec:CP}.

The M-step updates for $\bsbeta$ and $\bsR$ in the VP and CP models appear in Equations (\ref{eq:beta}) and (\ref{eq:mstepr}), given the appropriate definitions of $\bse$ and $\bsS$. The estimates for $\Gamma_g$ appearing in Equation (\ref{eq:gammag}) apply as well, except in this case the $\Gamma_g$ are all scalars.

The VP model estimates the persistence parameters $\alpha_{uv}$, whereas the CP model fixes them at 1. Let $\partial \bsS_*/\partial \alpha_{gt} =\bds{\Delta}^{gt}$. For example, when each student is linked to only one teacher in each year, $\bds{\Delta}^{gt}$ will be a sparse matrix with 1's in rows corresponding to year-$g$ observations, under columns corresponding to year-$t$ teachers. The score function for $\alpha_{gt}$ in the VP model is
\begin{equation*}
U(\alpha_{gt})=(\bsy^{\prime}-\bsbeta^{\prime}\bsX^{\prime})\bsR^{-1}\bds{\Delta}^{gt}\widetilde{\bse}-\te{tr}\left[\bsS_*^{\prime}\bsR^{-1}\bds{\Delta}^{gt}\left(\widetilde{\bsv}+\widetilde{\bse}\widetilde{\bse}^{\prime}\right)\right]
\end{equation*}
With the conditional moments and the other parameter blocks held fixed, $U(\bds{\alpha})$ is linear in the free persistence parameters $\bds{\alpha}=\{\alpha_{gt}:1\leq t<g\leq T\}$. When the corresponding coefficient matrix is nonsingular, a single Newton step provides an exact solution for $U(\bds{\alpha})=\bds{0}$.

\subsection{E-step}\label{sec:estep}
The E-step updates for all of the discussed models are identical, using the appropriate definitions of $\bsS,\bsG,\bse$, and $\bsR$.
The M-step equations and the observed data score require the first two moments, $\widetilde{\bse}$ and $\widetilde{\bsv}$, of $f\left(\bse|\bsy;\moparm\right)$. Using the method of \citet{henderson50,henderson75}, the moments are obtained from the gradient and Hessian of $\log f(\bsy,\bse;\moparm)$ with respect to $\bse$. The resulting estimates are
\begin{align}
\widetilde{\bsv}&=\left(\bsS^{\prime}\bsR^{-1}\bsS+\bsG^{-1}\right)^{-1}\label{eq:varetahat}\\
\widetilde{\bse}&=\widetilde{\bsv}\bsS^{\prime}\bsR^{-1}(\bsy-\bsX\bsbeta)\label{eq:etahat}
\end{align}
The expression for the EBLUP in Equation (\ref{eq:etahat}) is equivalent, via a matrix identity, to the perhaps more familiar expression
\begin{equation}
\widetilde{\bse}=\bsG\bsS^{\prime}\bsV^{-1}(\bsy-\bsX\bsbeta)\label{eq:etahat2}
\end{equation}
However, from a computational standpoint, (\ref{eq:etahat}) is much more efficient than (\ref{eq:etahat2}) since it does not require calculation of the full marginal covariance matrix $\bsV$. The calculation of (\ref{eq:etahat}) is relatively fast despite the large dimension of $\bsR$ because both $\bsS^{\prime}$ and  $\bsR^{-1}$ are sparse.

\subsection{EM Standard Errors}\label{ssec:EMSE}
One criticism of the EM algorithm is that it does not produce the Hessian of the observed data log-likelihood at the MLE $\widehat{\moparm}$ as a byproduct. The work we have already done, however, makes it possible to compute the observed data information matrix directly without working through a correction to the complete-data information matrix, as done by \citet{louis}. By the identity following Equation~(\ref{eq:emscore1}), the M-step gradients give the observed score when the conditional moments are evaluated at the same parameter vector. We may calculate the observed information matrix,
\begin{align}\label{eq:hessian}
-\left.\partial S(\moparm)/\partial \moparm^{\prime} \right|_{\moparm=\widehat{\moparm}},
\end{align}
with a finite-difference approximation at the MLE $\widehat{\moparm}$. All conditional moments must be recomputed at every perturbed parameter vector; holding the E-step moments fixed differentiates $Q$ rather than the observed data log-likelihood. This method is suggested by \citet{jam00}, who propose using either a forward or central difference approximation, or a Richardson extrapolation \citep{lindfield}. Joint observed-information standard errors are obtained from the inverse of this full matrix.

It is also useful to calculate standard errors for the predicted random effects. With $\bsbeta$, $\bsG$, and $\bsR$ known, $\widetilde{\bsv}$ is the conditional covariance of $\bse$ and the mean squared prediction error matrix of its conditional mean. Estimating $\bsbeta$ adds uncertainty \citep{sasbook}. With $\bsG$ and $\bsR$ known and $\bsbeta$ estimated by generalized least squares, the prediction error covariance matrix of the BLUP appears in block $\bds{C}_{22}$ of the following matrix \citep{mclean}
\begin{displaymath}
\bds{C}=
\begin{pmatrix}
\bds{C}_{11}&\bds{C}_{12}\\
\bds{C}_{21}&\bds{C}_{22}
\end{pmatrix}
=
\begin{pmatrix}
\bsX^{\prime}\bsR^{-1}\bsX&\bsX^{\prime}\bsR^{-1}\bsS\\
\bsS^{\prime}\bsR^{-1}\bsX&\bsS^{\prime}\bsR^{-1}\bsS+\bsG^{-1}
\end{pmatrix}^{-1}.
\end{displaymath}
Substituting fitted covariance parameters gives a plug-in prediction error covariance; it does not account for their estimation uncertainty.
The block $\bds{C}_{11}=(\bsX^{\prime}\bsV^{-1}\bsX)^{-1}$ gives the usual plug-in covariance matrix for $\widehat{\bsbeta}$. It need not equal the $\bsbeta$ block of the inverse full joint observed information matrix, because the observed cross-information between $\bsbeta$ and the covariance parameters need not vanish.

\subsection{Convergence and Initial Values of the EM Algorithm}
The EM algorithm converges to a stationary value of the observed data likelihood as long as $Q(\moparm;\moparm^{\prime})$ is continuous in both $\moparm$ and $\moparm^{\prime}$, and the parameter space is compact \citep{wu83}. Although the parameter space for $\moparm$ is not compact for Model (\ref{eq:LMM}), this regularity condition can be satisfied by a truncation of the parameter space \citep{mcculloch94, demidenko}.

One possible convergence criterion is to stop the algorithm when the relative change in the log-likelihood at iteration $k$, $l(\moparm^{(k)})$, is less than a fixed tolerance,
\[\frac{|l(\moparm^{(k)})-l(\moparm^{(k-1)})|}{\max\{1,|l(\moparm^{(k)})|\}} < w.\]
In general, we use $w=10^{-7}$. At an interior solution, a near-zero observed score together with a negative definite observed log-likelihood Hessian provides a local numerical convergence check. As with any iterative maximization routine, there is no way to guarantee that the EM algorithm will converge to the global maximum of the likelihood, given a single set of initial values $\moparm_0$. It is advisable to compare the results of the algorithm after starting from different sets of initial values. For the VAMs in Section \ref{sec:gp}, we did not find any sensitivity to the choice of $\moparm_0$.

\subsection{Restricted Maximum Likelihood}
The EM algorithm also extends to restricted maximum likelihood (REML) estimation.
For covariance updates with $\bsS$ held fixed, REML uses prediction-error
blocks of $\bds{C}$ from Section~\ref{ssec:EMSE}, evaluated at the current
covariance parameters, with $\bsbeta$ fitted by generalized least squares.
The random-effect adjustment uses $\bds{C}_{22}$ in place of
$\widetilde{\bsv}$; residual adjustments use
$[\bsX\ \bsS]\bds{C}[\bsX\ \bsS]^{\prime}$, including the fixed-effect
and cross blocks. These adjustments account for estimating $\bsbeta$.
\citet{karl2026reml} illustrates this distinction for simple variance
components. In VP, $\bsS_*$ replaces $\bsS$; it depends on the persistence parameters,
whose updates also require $\bds{C}_{12}$.
The original numerical results reported here remain ML results.

\section{Implementation of the EM Algorithm}\label{sec:implement}
We have implemented estimation of the GP, VP and CP models in the R \citep{R} package GPvam \citep{gpvam} for educational value-added assessments as an example of applying the proposed EM algorithm. Our program takes advantage of the sparseness of the design and certain covariance matrices, and handles large data sets relatively well. Because the program was custom-designed for these VAMs, it requires minimal input. The user must supply a data frame with columns for test scores, year of observation, student ID, and teacher ID. Optionally, other columns may be included for additional covariates in the $\bsX$ matrix; these are declared to the program through an R \texttt{formula} statement. Sparse matrices are constructed and handled via the R package Matrix \citep{Matrix}.

The GP model requires specification of a complex random effects structure.  \citet{doran} provide a tutorial to the implementation of VAMs in R using the functions \texttt{lme} and \texttt{lmer}. However, \citet{lme} explain that, for a less complicated multi-membership model, data sets with more than 200 teachers require several tricks to program with \texttt{lme}, and often fail to converge. 
GPvam automatically builds the sparse design matrix for the random effects, and performs well in the application in Section \ref{sec:data} which contains 4781 teacher effects for GP.R.

Although GPvam has been tailored to the estimation of VAMs, the R code may be generalized to other applications involving linear mixed models. New code would need to be written to build the application-specific $\bsS$ and $\bsG$ matrices but iterative parts of the program  would not need to be adjusted. The EM algorithm may also be extended for efficient estimation of non-nested, nonlinear mixed models. The use of a nonlinear link function will require an integral approximation in the E step.
It would also be possible to impose structure on the $\bsR$ matrix, such as autoregressive, compound symmetric, or Toeplitz. The procedures for obtaining the score functions of the parameters of an unstructured $\bsR$ may serve as a template for these other situations.

As mentioned in Section \ref{ssec:mg}, the M-step update for $\bsR$ in GP.R, rGP.R, and VP requires extra computational work due to the lack of a readily-available closed form solution. We use a Newton-Raphson algorithm to calculate the M-step update. The standard NR algorithm for solving $U(\bsR)=\bds{0}$ often diverges when the initial $\bsR_0$ is too far from the maximum. To improve the stability of the routine, we modify the appropriate Hessian by adding a scaled diagonal matrix during the first few M-step updates for $\bsR$. This results in a hybrid of a Newton and a gradient descent method that produces more reliable convergence when the initial value is far away from the critical point \citep{numopt}.

To compare the performance of GPvam and SAS (with the EFFECT statement of PROC GLIMMIX) in implementing the models presented in Section \ref{sec:gp}, we consider a data set with 6236 observations on 2834 students over 3 years, with 102, 104, and 98 teachers in each year, respectively. Table \ref{tab:sas} gives the results. GP.R and rGP.R each failed in SAS after encountering a negative-definite covariance matrix, while GP.G ran out of memory in SAS after a few minutes. The application in Section \ref{sec:data} involves a much larger data set than the one used in this example.

\begin{table}
\caption{Run times in minutes}
\centering
\label{tab:sas}
\begin{tabular}{lll}
\toprule
Model&GPvam&SAS\\
\midrule
CP&4.2&55\\
VP&5.5&N.A.\\
GP.G&12&Failed\\
rGP.R&80&Failed\\
GP.R&114&Failed\\
\bottomrule
\end{tabular}
\end{table}

\section{Application}\label{sec:data}
We apply the models to the data set analyzed by \citet{mariano10}, which is available in the supplementary material of \citet{mc10}. According to \citet{mc10}, the data come from vertically linked mathematics standardized test scores from grades 1--5 for a cohort of students from a large urban US school district.

The data have been pre-processed by \citet{mc10}, and we further process the data by removing observations with no student link, as well as observations missing both the test score and the teacher link. The resulting data set consists of 26019 observations on 9295 students over 5 years. For grades 1 through 5, there are 338, 318, 306, 321, and 259 teachers, respectively. This results in a total of 4781 teacher effects for GP.R.	The data set does not contain any additional covariates, so the fixed effects modeled include  a mean for each year.

\begin{figure}
\caption{Estimated $\bsG$ and $\bsR$ matrices from GP.R. The covariance matrix is on the left, and the correlation matrix is on the right.}
$\bsR$:
\label{GPparm}
\begin{flalign*}
&\begin{pmatrix}
0.741 &0.478 &0.463 &0.456 &0.392\\
0.478 &0.705 &0.523 &0.516 &0.449\\
0.463 &0.523 &0.736 &0.563 &0.484\\
0.456 &0.516 &0.563 &0.688 &0.509\\
0.392 &0.449 &0.484 &0.509 &0.565\\
\end{pmatrix}
\begin{pmatrix}
1.000 &0.661 &0.626 &0.639 &0.606\\
0.661 &1.000 &0.726 &0.740 &0.711\\
0.626 &0.726 &1.000 &0.791 &0.750\\
0.639 &0.740 &0.791 &1.000 &0.817\\
0.606 &0.711 &0.750 &0.817 &1.000
\end{pmatrix}&
\end{flalign*}
$\bsGamma_1$:
\begin{flalign*}
&\begin{pmatrix}
0.443 &0.121 &0.120 &0.107 &0.095\\
0.121 &0.100 &0.088 &0.084 &0.077\\
0.120 &0.088 &0.087 &0.083 &0.076\\
0.107 &0.084 &0.083 &0.080 &0.074\\
0.095 &0.077 &0.076 &0.074 &0.069
\end{pmatrix}
\begin{pmatrix}
1.000 &0.575 &0.610 &0.568 &0.541\\
0.575 &1.000 &0.941 &0.941 &0.920\\
0.610 &0.941 &1.000 &0.994 &0.986\\
0.568 &0.941 &0.994 &1.000 &0.995\\
0.541 &0.920 &0.986 &0.995 &1.000
\end{pmatrix}&
\end{flalign*}
$\bsGamma_2$:
\begin{flalign*}
&\begin{pmatrix}
0.281&0.059&0.039&0.042\\
0.059&0.025&0.023&0.020\\
0.039&0.023&0.024&0.020\\
0.042&0.020&0.020&0.017
\end{pmatrix}
\begin{pmatrix}
1.000&0.703&0.478&0.593\\
0.703&1.000&0.951&0.967\\
0.478&0.951&1.000&0.979\\
0.593&0.967&0.979&1.000
\end{pmatrix}&
\end{flalign*}
$\bsGamma_3$:
\begin{flalign*}
&\begin{pmatrix}
0.248&0.032&0.024\\
0.032&0.015&0.015\\
0.024&0.015&0.015
\end{pmatrix}
\begin{pmatrix}
1.000&0.516&0.394\\
0.516&1.000&0.979\\
0.394&0.979&1.000
\end{pmatrix}&
\end{flalign*}
$\bsGamma_4$:
\begin{flalign*}
&\begin{pmatrix}
0.130&0.038\\
0.038&0.030
\end{pmatrix}
\begin{pmatrix}
1.000&0.612\\
0.612&1.000
\end{pmatrix}&
\end{flalign*}
$\bsGamma_5:$ $0.146$\newline
\end{figure}

\begin{figure}
\caption{Standard errors for current year teacher ratings for year-3 teachers.}
\label{ratings3_SE}
\centering
\includegraphics[scale=.45]{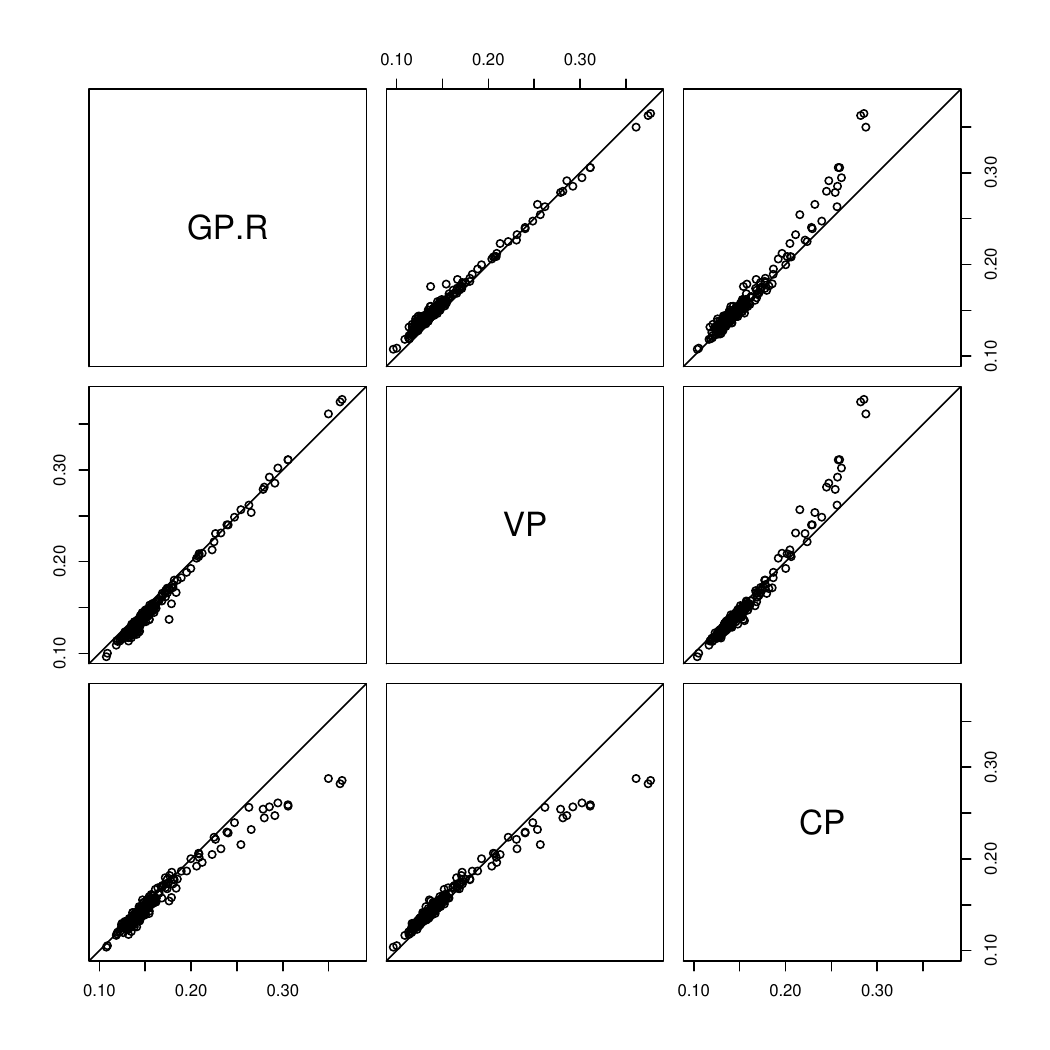}
\end{figure}
\begin{table}[htbp]
  \centering
  \caption{Estimates for yearly means from GP.R}
    \begin{tabular}{rrr}
    \addlinespace
    \toprule
          & Estimate & Std. Error \\
    \midrule
    Year 1 & 3.395 & 0.030 \\
    Year 2 & 3.996 & 0.029 \\
    Year 3 & 4.726 & 0.023 \\
    Year 4 & 5.309 & 0.022 \\
    Year 5 & 5.984 & 0.025 \\
    \bottomrule
    \end{tabular}%
  \label{tab:yearmean}%
\end{table}%

\begin{table}[htbp]
  \centering
  \caption{Persistence Parameters from VP}
    \begin{tabular}{rrr}
    \addlinespace
    \toprule
		&Estimate&S.E.\\
		\midrule
    $\alpha_{21}$ & 0.18  & 0.02 \\
    $\alpha_{31}$ & 0.19  & 0.02 \\
    $\alpha_{41}$ & 0.17  & 0.02 \\
    $\alpha_{51}$ & 0.15  & 0.02 \\
    $\alpha_{32}$ & 0.22  & 0.02 \\
    $\alpha_{42}$ & 0.14  & 0.02 \\
    $\alpha_{52}$ & 0.16  & 0.02 \\
    $\alpha_{43}$ & 0.13  & 0.02 \\
    $\alpha_{53}$ & 0.09  & 0.02 \\
    $\alpha_{54}$ & 0.29  & 0.03 \\
    \bottomrule
    \end{tabular}%
  \label{tab:pparms}%
\end{table}%

We fit each of the models GP.R, rGP.R, GP.G, VP, and CP to this data set using the program GPvam. Table \ref{tab:yearmean} lists the estimated yearly means from GP.R: the results from the other models are similar. Figure \ref{GPparm} gives the maximum likelihood estimates of the covariance parameters from GP.R. Models rGP.R and GP.G are valid for this data set because the scores from each year are on the same scale of measurement. The estimates of current-year teacher effects are nearly identical for the three variations of the GP model, with correlations of 0.998 or higher among the estimated effects. The agreement between GP.R and rGP.R is not surprising given the extremely high correlations between future year effects seen in the $\bsGamma_g$ matrices of Figure \ref{GPparm}. With a simplified covariance structure, rGP.R converges after 48 iterations in 7\% of the time it takes the 431 iterations needed for GP.R to converge. For this data set, the reduced model rGP.R appears to provide a good alternative. Slow convergence in the neighborhood of a maximum that lies near the boundary of the parameter space is a well-known property of the EM algorithm \citep{demidenko}. However, the time used by the EM algorithm is worthwhile, because the faster NR algorithms are prone to failure in these settings.

Fitting the CP and VP models with the EM algorithm results in approximately the same correlations among the GP.R, VP, and CP estimates as found by \citet{lock07} and \citet{mariano10}. The persistence parameters for the VP model in Table \ref{tab:pparms} are similar to those provided by \citet{mc10}, who modeled students with random effects. The persistence parameters are all significantly different from 1, indicating that the assumption of complete persistence is not compatible with this data set.

Using the EM algorithm, we obtain correlation patterns for GP.R that are similar to those in Figures 2 and 3 of \citet{mariano10}. However, we note that \citet{mariano10} obtained these results after careful choice of an informative prior that allowed for strong correlations between future year effects. In simulation studies, they found that a minimally informative Wishart prior for covariance parameters could result in credible intervals for the correlations that did not include the true values. The EM algorithm gives maximum likelihood estimates that do not need any specifications of prior distributions.

Figure \ref{ratings3_SE} compares the standard errors associated with the predicted teacher effects for GP.R, VP, and CP. As stated in Section~\ref{ssec:EMSE}, the standard errors are calculated as a by-product of the EM algorithm. The values for the larger standard errors, which likely correspond to teachers with relatively fewer observations, are inflated when moving from the CP to the VP or GP models. This is interesting because the prediction intervals are used by some researchers to classify teachers as below-average, average, or above average \citep{draper95,lock07}. Despite the inflation of some of the standard errors seen when moving from the CP to the VP model, most of the prediction errors are smaller in the VP model. An advantage of maximum likelihood estimation is that the standard errors for the teacher effects, derived in Section \ref{ssec:EMSE}, are free from the influence of potentially informative prior distributions.

The models fit to this data set have a number of assumptions that were stated in Section~\ref{sec:gp}. All models assume that the teacher effects in $\bse$ and the residual student effects in $\bds{\epsilon}$ are independent.
If data are collected in a designed experiment, with students randomly assigned to teachers, this assumption is reasonable.
Most data used in VAMs are observational, though, so this assumption is violated if, say, some teachers are regularly
assigned the best students. In that case, the  effects ascribed to teachers may actually be more
properly attributed to the students who take those teachers. The models are also assumed to be ``correct'' in that they are assumed
to include all factors relevant to the response. The data set analyzed here did not contain information on student-level
covariates such as socioeconomic status, for example, and it is possible that including such covariates in the model would 
change the estimated teacher effects. 
These models further assume that missing test scores are missing at random. Conditional on observed scores, covariates, and recorded teacher history, the probability of a missing-data pattern does not depend on unobserved test scores. Dependence on a student's latent ability beyond what the observed data reflect would generally make the missingness nonignorable.
Finally, although the multiresponse models considered
here relate later test scores to earlier test scores through the within-student correlations, the models imply that,
conditionally on $y_{i,g-1}$, the  relationship between $y_{ig}$ and $y_{i,g-1}$ is linear. 

The usefulness of VAM scores for measuring teacher effectiveness depends on the quality of the tests as
 measures of student achievement \citep{koretz}, and many aspects of teacher contributions may be unrelated
by standardized tests \citep{braun}.
A relevant discussion of the utility and limitations of multilevel models is given by \citet{draper95}, who urges a careful examination of the nature of the sampling in the study.
Thus, caution is needed when interpreting a teacher's VAM score.

With these limitations in mind, VAMs can provide valuable information for improving the educational system \citep{harris10}.
In this example, the GP model indicates that the current-year teacher has the highest effect on student scores, and that the
current-year effects are, on average, stronger in earlier grades than in later grades. The estimates of relative sources
of variability also provide valuable information: in grade 1, current-year teachers (or other classroom effects that are associated with teachers) account for approximately 36\% of the variability
in student scores but that percentage drops to 16--20\% in grades 4 and 5. \citet{ballou} and \citet{lock07} note that teacher effects from the first year are most susceptible to bias resulting from nonrandom student assignment to classrooms.

\section{Conclusions}
In this article, we have developed a method for computing maximum likelihood estimates for a class of multiple membership models
in which lower-level units progress through sequential higher-level units  \citep{mariano10}. The EM algorithm offers an efficient method of computation, taking advantage of matrix sparsity and requiring inversion of a matrix whose dimension depends on the number of random effects, rather than on the total number of observations as in other implementations. The algorithm produces stable behavior even when the covariance matrix for the random effects is nearly singular. We have implemented the proposed methods in the R package GPvam. The availability of maximum-likelihood estimates should be useful for those preferring Bayesian estimation as well, providing a sensitivity analysis to their choice of priors. We hope that this user-friendly implementation of the model will facilitate further empirical study of the model's properties.

In the educational context, the GP and other models fit provide a great deal of flexibility for studying relative effects of teachers
on their students' current and future achievement.
In some cases the full flexibility of the GP model may be needed to summarize the data structure;
in others, being able to examine the estimates from the GP model may show that a simpler structure adequately describes  the data.
Our algorithm readily calculates standard errors for the predicted teacher effects (VAM scores).
In many applications, the standard errors of the teacher effects are quite large \citep[p.~45]{braun},
so that including the standard errors along with the point estimates can help distinguish ``real''
effects from random variation.

Although the computational methods and software were developed in the educational setting, they can be used in many other applications as well, substituting the lower-level units for ``students'' and the higher-level units for ``teachers''.
Similar models have been considered for studying the relative contributions of health care professionals or clinics to 
patient outcomes \citep{zaslavsky}, and the models can be applied in any setting where different higher-level units
sequentially affect the outcomes of lower-level units.

The algorithm presented here and the code in package GPvam may be extended to other multiple membership models.
One extension would be to allow an expanded covariance structure in which the same higher-level unit may be
associated with multiple responses.
This would allow the model to better fit situations in which a student had the same teacher for more than one year,
or in which a patient returned to a previous doctor.
This extension would require careful bookkeeping to track which doctors are repeats, but the basic M-step and E-step
of the algorithm would remain the same.
Our models and applications used only intercepts for the random effects, but the implementation may be extended
to include random slopes as well, and other more general multiple membership models described in \citet{browne01}.

 \appendix
 \section*{Appendix}
We show that the EM algorithm produces positive definite (PD) $\bsG$ matrix after each iteration. This assumes that the $\bsR$ matrix is PD after each iteration, which is true for the GP.G model specification in Section \ref{sec:alt}, and can be demonstrated for GP.R from Section \ref{ssec:gp} in the absence of incomplete data where the M-step update for $\bsR$ has an easily-obtainable solution. However, even in the presence of missing data, $\bsR$ will usually not be near the boundary of the parameter space, since the error variances on the diagonal of $\bsR$ will be positive as long as the model does not fit the data perfectly, and the intra-student effects are not likely to be perfectly correlated.

A much more common problem in estimating mixed models occurs when the estimated $\bsG$ matrix is not PD \citep[Section 5.6.1]{verb99}. This may happen when 0 variance components are estimated, or when random effects are perfectly correlated, the latter being a significant concern for the future teacher effects of the GP VAM. Notice that $\bsG$ is a block-diagonal portion of $\widetilde{\bds{v}}+\widetilde{\bse}\widetilde{\bse}^{\prime}$. The matrix $\widetilde{\bds{v}}$ is defined by Equation~(\ref{eq:varetahat}). Thus, $\widetilde{\bds{v}}$ is PD as long as the initial $\bsG$ is PD, because $\bsS^{\prime}\bsR^{-1}\bsS$ is positive semi-definite. Furthermore, $\widetilde{\bse}\widetilde{\bse}^{\prime}$ is positive semi-definite, and the sum of a PD and a positive semi-definite matrix is PD.
\section*{Acknowledgments}
This research was partially supported by the National Science Foundation under grant DRL-0909630. Any opinions, findings, and conclusions or recommendations expressed in this material are those of the author and do not reflect the views of the National Science Foundation or Arizona State University.

\bibliographystyle{asabst}
\bibliography{disbib}

\end{document}